# Robust Watermarking in Multiresolution Walsh-Hadamard Transform

Gaurav Bhatnagar[1] and Balasubramanian Raman[2]
Department of Mathematics, Indian Institute of Technology Roorkee, Roorkee-247 667, India
[1]goravdma@gmail.com, [2]balaiitr@ieee.org

*Abstract*—In this paper, a newer version of Walsh-Hadamard Transform namely multiresolution Walsh-Hadamard Transform (MR-WHT) is proposed for images. Further, a robust watermarking scheme is proposed for copyright protection using MR-WHT and singular value decomposition. The core idea of the proposed scheme is to decompose an image using MR-WHT and then middle singular values of high frequency sub-band at the coarsest and the finest level are modified with the singular values of the watermark. Finally, a reliable watermark extraction scheme is developed for the extraction of the watermark from the distorted image. The experimental results show better visual imperceptibility and resiliency of the proposed scheme against intentional or un-intentional variety of attacks.

## I. INTRODUCTION

Now a days, some very crucial issues of digital media are duplication, distribution, editing, copyright protection etc. The main reason of these kind of issues is development of internet and multimedia technology. As a solution, Digital Watermarking is used very frequently. Hence, digital watermarking becomes very attractive research topic and many many taxonomies for digital watermarking have been proposed. Among these, the most common taxonomies are embedding in spatial[1] and transform domain. Watermarking schemes of transform domain, such as discrete Fourier transform (DFT)[2], discrete cosine transform(DCT)[3], [4], Walsh-Hadamard Transform[9] and discrete wavelet transform(DWT)[6], [7], [8], Multi-Polarity Hadamard transform[9] are more popular, since they provide more advantages and better performances than those in the spatial domain. The wavelet-based methods become more prevalent in transform-domain watermarking algorithms due to their excellent spatial localization, frequency spread and multiresolution characteristics. In the recent years, singular value decomposition is used as a new tool for watermarking[10], [11], [12]. The most common approach is to modify the singular values of the host image by singular values of the watermark. These modified singular values are combined with the known component to get the watermarked image.

In this paper, a robust algorithm of watermarking is introduced for images. The proposed watermarking scheme is based on multiresolution Walsh-Hadamard Transform(MR-WHT) and SVD approaches. For embedding, the host image is transformed into frequency domain using MR-WHT and then watermark is embedded in the middle singular values of the high frequency sub-bands at the coarsest and the finest level. The largest singular values are more important to the image quality and the smallest singular values are more sensitive to the noise. Here the middle singular values are selected to embed the watermark. Further, an efficient watermarking extraction scheme is introduced for finding the estimate of the watermark from both the coarsest and the finest level high frequency sub-bands.

This paper is organized as follows: In section II and III introduction to the Multiresolution Walsh-Hadamard Transform and singular value decomposition. Our proposed watermarking embedding and extraction schemes are described in section IV. Section V presents experimental results using proposed watermarking scheme and finally the concluding remarks are given in section VI.

## II. MULTIRESOLUTION WALSH-HADAMARD TRANSFORM (MR-WHT)

Unlike Fourier and Cosine Transforms, in the case of Walsh-Hadamard Transform the basic functions are not sinusoids. The basic functions are based on square or rectangular waves with peaks of $\pm 1$. Here the term rectangular wave refers to any function of this form, where width of the pulse may vary. One primary advantage of the transform is that the computations are very simple. When we project an image onto the basis functions, all we need to do is multiply each pixel by 1. Mathematically, it is defined as:

$$WH(u) = \frac{1}{N} \sum_{x=0}^{N-1} f(x) \, (-1)^{\sum_{i=0}^{n-1} b_i(x) \, b_i(u)} \qquad (1)$$

where $N = 2^n$ and $b_k(z)$ is the $k^{th}$ bit in the binary representation of $z$. For example, if $n = 3$ and $z = 6$ ($\approx 110$ in binary) then $b_0(z) = 0$, $b_1(z) = 1$ and $b_2(z) = 1$. As inverse Fourier and Cosine transform, inverse Walsh-Hadamard transform is defined as:

$$f(x) = \sum_{u=0}^{N-1} WH(u) \, (-1)^{\sum_{i=0}^{n-1} b_i(x) \, b_i(u)} \qquad (2)$$

Due to separability of the transform, two dimensional WHT can be obtained by successively taking one dimensional WHT along both the axis. Hence, the WHT of 2D function $f(x,y)$ is written as

$$WH(u,v) = WHT^{y \to v}\{WHT^{x \to u}\{f(x,y)\}\} \qquad (3)$$

By definition, it is clear that WHT is not classified as a frequency transform, because the basis functions do not exhibit the frequency concept in the manner of sinusoidal functions.

However, it considers the number of zero crossings (or sign changes).

The multiresolution Walsh-Hadamard transform (MR-WHT) for an image is the modification of the Walsh-Hadamard transform (WHT) and is implemented in three steps:

*Step 1:* WHT is performed on image $F$, which is denoted by $\mathcal{F}$.

*Step 2:* Row wise modification results

$$\mathcal{F}_1(:,x) = \left\lfloor \frac{\mathcal{F}(:,2x-1)+\mathcal{F}(:,2x)}{2} \right\rfloor \quad (4)$$

$$\mathcal{F}_1\left(:,\frac{Width}{2}+x\right) = \mathcal{F}(:,2x-1)-\mathcal{F}(:,2x) \quad (5)$$

*Step 3:* Column wise modification results

$$\mathcal{F}_2(y,:) = \left\lfloor \frac{\mathcal{F}_1(2y-1,:)+\mathcal{F}_1(2y,:)}{2} \right\rfloor \quad (6)$$

$$\mathcal{F}_2\left(\frac{Height}{2}+y,:\right) = \mathcal{F}_1(2y-1,:)-\mathcal{F}_1(2y,:) \quad (7)$$

where $F$, $\mathcal{F}$, $\mathcal{F}_1$ and $\mathcal{F}_2$ refer to original, WHT transformed, temporary and final MR-WHT images respectively. Figure 1 shows the MR-WHT decomposition process of an image. Similarly, inverse MR-WHT is also implemented in three steps:

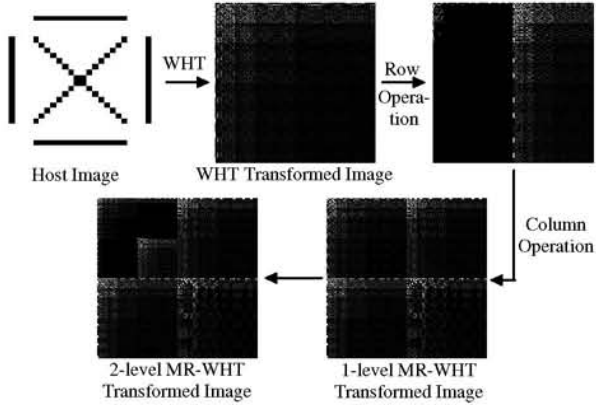

Fig. 1.  Multiresolution Walsh-Hadamard Transform of an Image

*Step 1:* Undo Column wise modification, that results

$$\mathcal{F}_2'(2y-1,:) = \mathcal{F}_2(y,:) + \left\lfloor \frac{\mathcal{F}_2\left(\frac{Height}{2}+y,:\right)+1}{2} \right\rfloor \quad (8)$$

$$\mathcal{F}_2'(2y,:) = \mathcal{F}_2'(2y-1,:) - \mathcal{F}_2\left(\frac{Height}{2}+y,:\right) \quad (9)$$

*Step 2:* Undo Row wise Modification, that results

$$\mathcal{F}_1'(:,2x-1) = \mathcal{F}_2'(:,x) + \left\lfloor \frac{\mathcal{F}_2'\left(:,\frac{Width}{2}+x\right)+1}{2} \right\rfloor \quad (10)$$

$$\mathcal{F}_1'(:,2x) = \mathcal{F}_1'(:,2x-1) - \mathcal{F}_2'\left(:,\frac{Width}{2}+x\right) \quad (11)$$

*Step 3:* Inverse WHT is performed on $\mathcal{F}_1'$ to get the reconstructed image $F'$.

where $\mathcal{F}_2$, $\mathcal{F}_2'$, $\mathcal{F}_1'$ and $F'$ refer to MR-WHT, temporary, reconstructed WHT and the original images respectively.

## III. SINGULAR VALUE DECOMPOSITION

Recently, Singular value decomposition is used very frequently in digital image processing due to excellent resist power of singular values i.e. singular values of an image are less affected if general image processing is performed. Moreover, the size of the matrices from SVD transformation is not fixed. It can be a square or a rectangle. Finally, singular values contain intrinsic algebraic image properties.

Let $A$ be a general real matrix or image of order $m \times n$. The singular value decomposition (SVD) of $A$ is the factorization

$$A = U * S * V^T \quad (12)$$

where $U$ and $V$ are *orthogonal* matrices of size $m \times m$ and $n \times n$ respectively while $S = diag(\sigma_1, \sigma_2, ..., \sigma_r)$, where $\sigma_i$, $i = 1(1)r$ are the singular values of the matrix $A$ with $r = min(m,n)$ and satisfying $\sigma_1 \geq \sigma_2 \geq ... \geq \sigma_r$. The first $r$ columns of $V$ are called *right singular vectors* and the first $r$ columns of $U$ are called *left singular vectors*. Here each singular value specifies the luminance of the image layer while the corresponding pair of singular vectors specify the geometry of the image layer.

## IV. PROPOSED ALGORITHM

Without loss of generality, let us assume that $F$ represents the host image of size $M_F \times M_F$, $W$ represents the initial watermark of size $M_W \times N_W$ and the watermark image is smaller than the host image by a factor $2^{Q_1}$ and $2^{Q_2}$ along both the axis, where $Q_1$ and $Q_2$ are any integers greater than or equal to 1. Since Walsh-Hadamard transform is applied on the square matrix (images), we have taken square image as host image however for rectangular images make those images square by adding rows/columns having elements zero. Watermark is embedded in HH sub-band at the coarsest and the finest level. The block diagram of proposed embedding and extraction scheme are shown in figures 2 and 3 respectively.

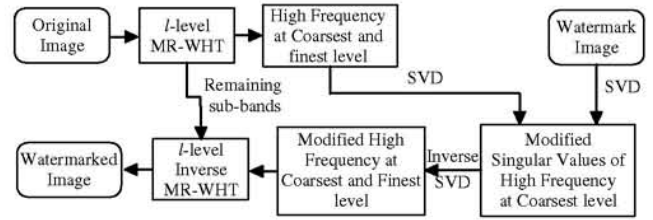

Fig. 2.  Watermark Embedding Algorithm

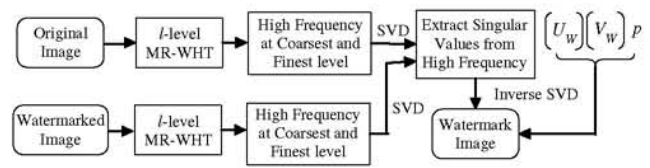

Fig. 3.  Watermark Extraction Algorithm

## A. Embedding Algorithm

The watermark embedding algorithm is formulated as follows:

1) Perform $L$-level MR-WHT on the host image which is denoted by $f_l^\theta$, where $\theta \in \{$ LL, LH, HL, HH $\}$ and $l \in [1, L]$.
2) Select HH sub-band from the coarsest and the finest level which is denoted by $f_\phi^{HH}$, where $\phi \in \{1, L\}$. Here 1 and $L$ stand for the coarsest and the finest level respectively.
3) The process of embedding is described as follows:
   - Apply SVD on both HH sub-band and watermark image,
   $$f_\phi^{HH} = U_{f_\phi^{HH}} \ S_{f_\phi^{HH}} \ V_{f_\phi^{HH}}^T \quad (13)$$
   $$W = U_W \ S_W \ V_W^T \quad (14)$$
   - Modify the middle singular values of the HH sub-band as follows:
   $$\sigma_{f_\phi^{HH}}^{new}(i+p_\phi) = \sigma_{f_\phi^{HH}}(i+p_\phi) + \frac{\alpha * \sigma_W(i)}{max(\sigma_{f_\phi^{HH}})} \quad (15)$$

   where $\alpha$ is the watermark strength, $i = 1, 2, 3, ..., min(M_W, N_W)$ and $p_\phi$ are the factors for choosing middle singular values from HH sub-band at the coarsest and the finest level. For example, if the value of $p_\phi$ is 63 then the watermark singular values are embedded among 64 to 63+$min(M_W, N_W)$ singular values of HH sub-band. Since, the largest singular values are more important to the image quality, and the smallest singular values are more sensitive to the noise, middle singular values are selected to embed the watermark.
   - Perform inverse SVD to construct the watermarked $f_\phi^{HH}$,
   $$\left(f_\phi^{HH}\right)^{new} = U_{f_\phi^{HH}} \ S_{f_\phi^{HH}}^{new} \ V_{f_\phi^{HH}}^T \quad (16)$$

4) Map modified sub-band to its original position and $L$-level inverse MR-WHT is performed to get the watermarked image.

## B. Extraction Algorithm

The objective of the watermark extraction is to obtain the estimate of the original watermark. For watermark extraction, host and watermarked images, $V_W$, $U_W$ and $p_\phi$ are required. The extraction process is formulated as follows:

1) Perform $L$-level MR-WHT on the host and watermarked images which are denoted by $f_l^\theta$, $fw_l^\theta$, where $\theta \in \{$ LL, LH, HL, HH $\}$ and $l \in [1, L]$.
2) Select HH sub-band from the coarsest and finest level from both images, which are denoted by $f_\phi^{HH}$ and $fw_\phi^{HH}$ respectively, where $\phi \in \{1, L\}$. Here 1 and $L$ stand for the coarsest and the finest level respectively.
3) In this step, sub-bands $fw_\phi^{HH}$ and $f_\phi^{HH}$ are used for extraction because watermark is embedded in HH sub-band at the coarsest and the finest level. Extraction process is described as follows:
   - Perform SVD on both $f_\phi^{HH}$ and $fw_\phi^{HH}$,
   $$f_\phi^{HH} = U_{f_\phi^{HH}} \ S_{f_\phi^{HH}} \ V_{f_\phi^{HH}}^T \quad (17)$$
   $$fw_\phi^{HH} = U_{fw_\phi^{HH}} \ S_{fw_\phi^{HH}} \ V_{fw_\phi^{HH}}^T \quad (18)$$
   - The estimate of singular values of watermark is given by
   $$\sigma_{W_\phi}^{ext}(i) = \frac{S_{fw_\phi^{HH}}(i+p_\phi) - S_{f_\phi^{HH}}(i+p_\phi)}{\alpha/max(\sigma_{f_\phi^{HH}})} \quad (19)$$
   - Perform inverse SVD to construct the extracted watermark,
   $$W_\phi^{ext} = U_W \ S_W^{ext} \ V_W^T \quad (20)$$

## V. RESULTS AND DISCUSSIONS

The robustness of proposed scheme is demonstrated using MATLAB. Different standard gray scale images of size $512 \times 512$ are used as the host images namely Payaso, Yacht and Zelda. For watermark, three gray scale images of size $64 \times 64$ namely Peacock, Cup and IEEE CS are used. Peacock, Duck and IEEE CS are embedded into Payaso, Yacht and Zelda images respectively. The watermarked image quality is measured using PSNR (Peak Signal to Noise Ratio). The factor for choosing middle singular values ($p_\phi$) is taken as 63 and 32 for coarsest and finest level respectively.

To verify the presence of watermark, the correlation coefficient between the original and the extracted singular values is given by

$$\rho(w, \bar{w}) = \frac{\sum_{i=1}^{r} (w(i) - w_{mean}) \ (\bar{w}(i) - \bar{w}_{mean})}{\sqrt{\sum_{i=1}^{r} (w(i) - w_{mean})^2} \sqrt{\sum_{i=1}^{r} (\bar{w}(i) - \bar{w}_{mean})^2}} \quad (21)$$

where $w$, $\bar{w}$, $w_{mean}$ and $\bar{w}_{mean}$ are the original, extracted singular values, mean of original and extracted singular values and $r = min(M_W, N_W)$. In figure 4, all the original and watermarked images are shown. Original and extracted watermarks are shown in figure 5. The corresponding PSNR and the correlation coefficient values are given in table I.

TABLE I
PEAK SIGNAL TO NOISE RATIO AND CORRELATION COEFFICIENTS OF ALL TEST IMAGES

| Image | PSNR | $\rho$ | |
|---|---|---|---|
| | | From Finest level | From Coarsest level |
| Payaso | 48.7463 | 0.9999 | 0.9959 |
| Yacht | 46.3570 | 1 | 0.9997 |
| Zelda | 47.8401 | 1 | 0.9983 |

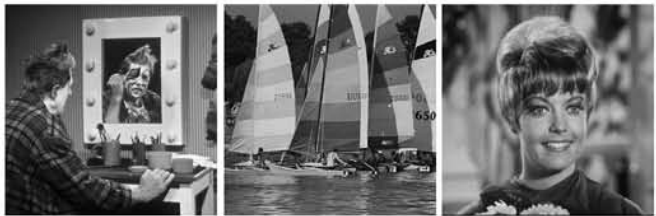

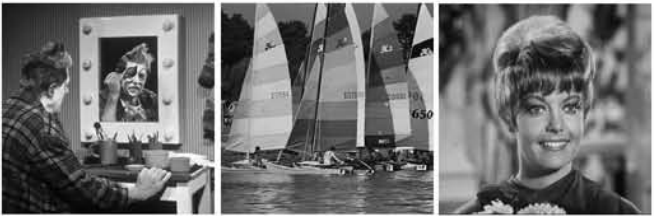

Fig. 4. a) Original b) Watermarked images.

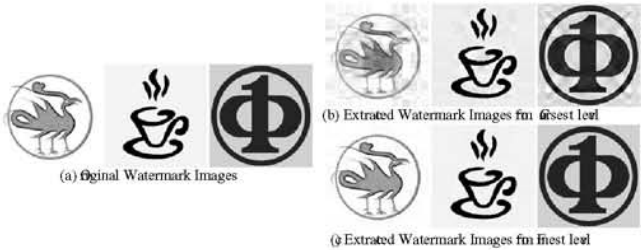

Fig. 5. a) Original b) Extracted watermark images.

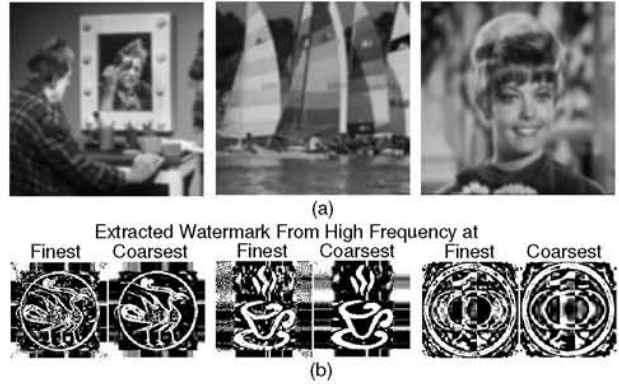

Fig. 6. Results for Gaussian Blurring

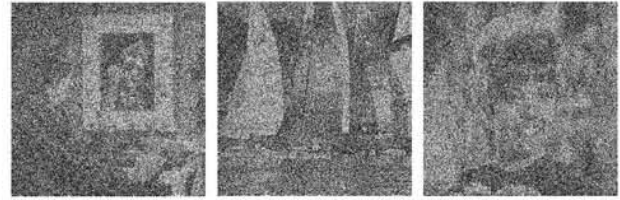

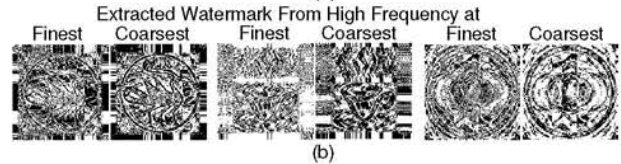

Fig. 7. Results for Gaussian Noise Addition (100%)

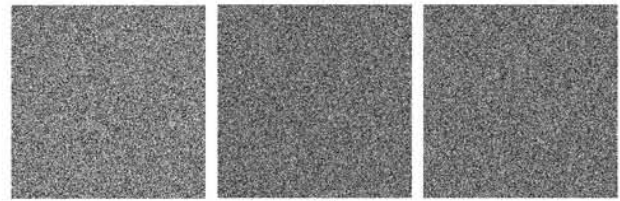

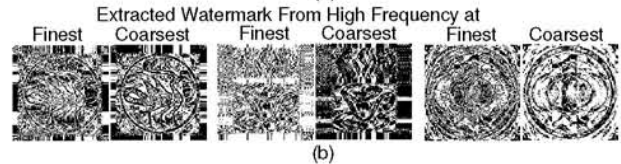

Fig. 8. Results for Salt & Paper Noise Addition (100%)

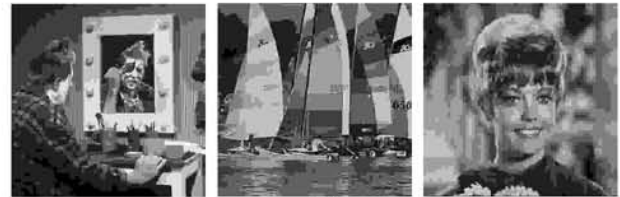

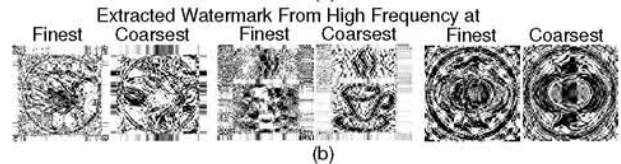

Fig. 9. Results for JPEG Compression (CR=100)

To investigate the robustness of the proposed algorithm, the watermarked image is attacked by Gaussian blurring, Gaussian and Salt & Paper noise addition, JPEG compression, Row and Column deletion, Pixelation, Cropping, Vertical Flipping, Horizontal flipping, Sharpen and Wrapping attacks. After these attacks on the watermarked image, the extracted gray scale watermark is compared with the original one. In table II, correlation coefficients are given for all extracted watermark images.

The most common manipulation in digital image is blurring. Watermark is extracted after applying $13 \times 13$ Gaussian blurring and the results are shown in the figure 6. To verify the robustness of the watermarking scheme, another measure is noise addition. In our experiments, $P\%$ additive Gaussian and Salt & Paper noise are added in the watermarked image. In figure 7 and 8, extracted watermarks from 100% Gaussian and Salt & Paper noise attacked watermarked images are shown. For the storage and transmission of digital data, image compression is often performed to reduce the memory and increase efficiency. Hence, we have also tested our algorithm for JPEG compression(CR 100) and the results are shown in figure 9. The proposed algorithm has also been tested for row-column deletion and resizing attacks. In row-column deletion, some rows and columns of the watermarked image are deleted randomly and then extract watermark. Figure 10 shows the result of randomly deleted 20 rows and 20 columns attack. Figure 11 shows the results of pixelation. Pixelation is the display of a digitized image where the individual pixels are apparent to a viewer. This can happen unintentionally when a low-resolution image designed for an ordinary computer display is projected on a large screen and each pixel becomes separately viewable. Another frequently used action on images is cropping. In figure 12, results for cropping are shown, when only 2.5% of area is remaining. Results for Vertical flipping, Horizontal flipping, sharpen and wrapping are shown in figures 13, 14, 15 and 16 respectively. Wrapping is the process of giving 3D effect to an object by wrap a selection around a shape. Figure 16 shows the extracted watermarks when object is wrapped around spherical shape.

TABLE II
CORRELATION COEFFICIENTS OF ALL EXTRACTED WATERMARK IMAGES AFTER ATTACKS

| Attacks | $\rho$ | | | | | |
|---|---|---|---|---|---|---|
| Image | Payaso | | Yatch | | Zelda | |
| | Watermark Extracted from | | | | | |
| | Finest | Coarsest | Finest | Coarsest | Finest | Coarsest |
| Gaussian Blur(13 × 13) | -0.0596 | -0.1698 | -0.1763 | -0.2369 | -0.1393 | -0.2704 |
| Gaussian Noise (100%) | 0.0398 | 0.0165 | 0.1461 | 0.2800 | 0.3477 | 0.1689 |
| Salt & Paper Noise (100%) | 0.0215 | 0.0403 | -0.0119 | 0.2469 | 0.3621 | 0.1154 |
| JPEG Compression (CR=100) | 0.5333 | 0.5635 | 0.5436 | 0.4265 | 0.4829 | 0.3106 |
| Row & Column Deletion (20 row & column) | -0.8604 | -0.9529 | -0.9354 | -0.9500 | -0.8269 | -0.9695 |
| Pixelation | 0.1771 | -0.0400 | -0.0719 | -0.1820 | 0.0083 | -0.1570 |
| Cropping (2.5% area remaining) | -0.9957 | -0.9687 | -0.9964 | -0.9969 | -0.9930 | -0.9923 |
| Vertical Flipping | -0.8441 | -0.1826 | -0.8746 | 0.9534 | 0.7827 | -0.0360 |
| Horizontal Flipping | -0.7861 | -0.9219 | 0.1804 | 0.8874 | 0.6571 | 0.8994 |
| Sharpen | -0.1998 | -0.3304 | -0.0659 | 0.2563 | 0.3092 | -0.2977 |
| Wrapping | 0.7894 | 0.8039 | -0.1158 | 0.7565 | -0.4442 | 0.5685 |

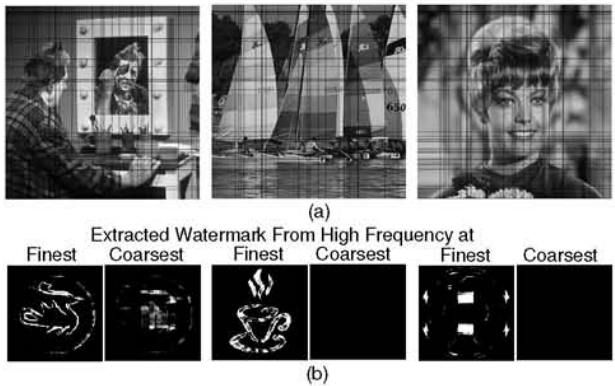

Fig. 10. Results for Row-Column Deletion (Randomly deleted 20 Rows and Columns)

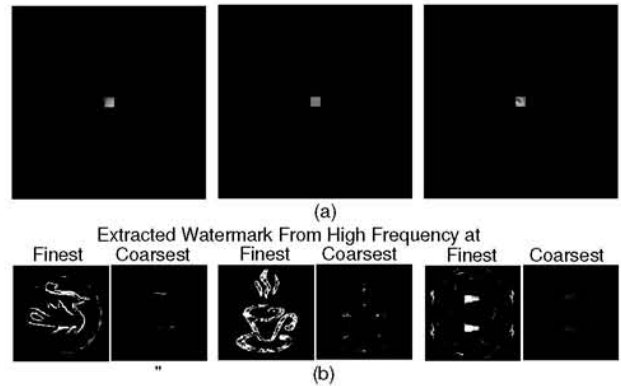

Fig. 12. Results for Cropping (2.5% of image area remaining)

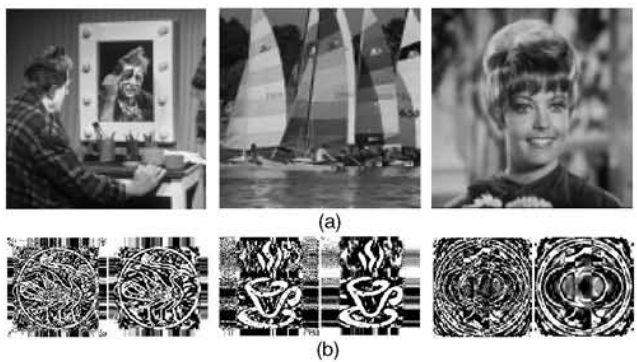

Fig. 11. Results for Pixelation

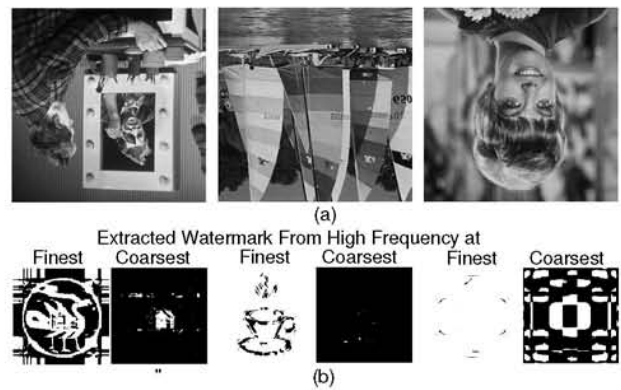

Fig. 13. Results for Vertical Flipping

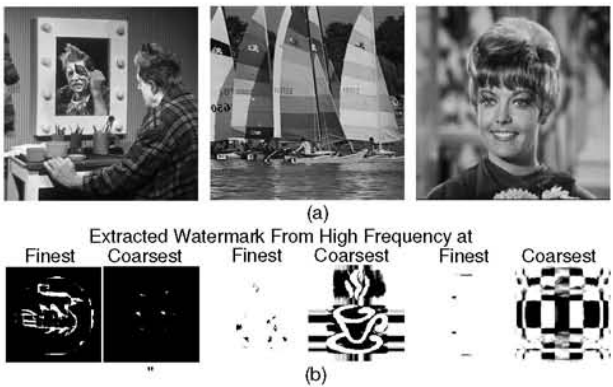

Fig. 14. Results for Horizontal Flipping

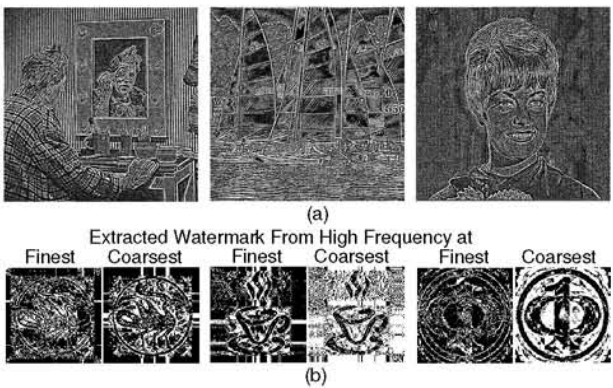

Fig. 15. Results for Sharpening (100%)

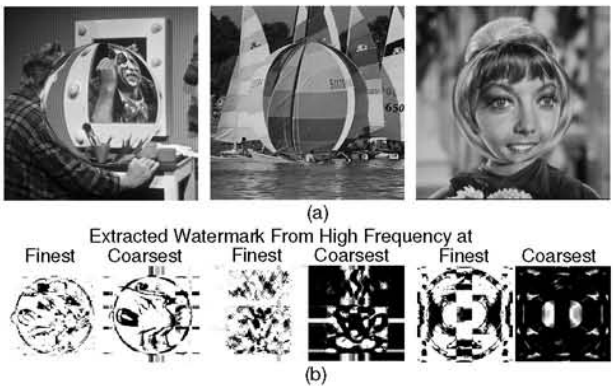

Fig. 16. Results for Wrapping (Wrap around a circle)

## VI. CONCLUSIONS

In this paper, multiresolution Walsh-Hadamard transform (for images) and its application in digital watermarking is proposed. Proposed algorithm is executed in the multiresolution Walsh-Hadamard domain using singular value decomposition (SVD). In which the watermark is a visually meaningful gray scale image/logo instead of a noise type Gaussian sequence. The factor for choosing middle singular values ($p_\phi$) are used as the keys for watermark extraction, which gives more complexity to the extraction process because without knowing these values no attacker can extract the data correctly. For the extraction of watermark, a reliable extraction scheme is constructed. Further, the proposed algorithm stands with various attacks, which shows the robustness of the proposed algorithm.

### ACKNOWLEDGEMENTS

One of the authors, Gaurav Bhatnagar, gratefully acknowledges the financial support of the Council of Scientific and Industrial Research, New Delhi, India through his Junior Research Fellowship (JRF) scheme (CSIR Award no.: 09/143(0559)/2006-EMR-I) for his research work.